\documentclass[aps,prl,10pt,twocolumn,superscriptaddress,showpacs,amsmath,amssymb,floatfix]{revtex4-1}

\usepackage{epsfig}
\usepackage{amsmath}
\usepackage{amssymb}
\usepackage{bm}
\usepackage{graphicx}
\usepackage{color}
\usepackage{graphicx,rotating}
\usepackage{txfonts}
\usepackage{microtype}
\usepackage{ulem}
\usepackage{epstopdf}

\begin{document}

%\title{Two-Color laser-polarization of positrons}
\title{Polarized positron beams via intense two-color laser pulses}

\author{Yue-Yue Chen}
\email{yue-yue.chen@mpi-hd.mpg.de}
\affiliation{Max-Planck-Institut f\"{u}r Kernphysik, Saupfercheckweg 1, 69117 Heidelberg, Germany}
\author{Pei-Lun He}
\affiliation{Max-Planck-Institut f\"{u}r Kernphysik, Saupfercheckweg 1, 69117 Heidelberg, Germany}
\affiliation{Key Laboratory for Laser Plasmas, Ministry of Education, and School of Physics and Astronomy, Shanghai Jiao Tong University, Shanghai 200240, China}
\author{Rashid Shaisultanov}
\affiliation{Max-Planck-Institut f\"{u}r Kernphysik, Saupfercheckweg 1, 69117 Heidelberg, Germany}
\author{Karen Z. Hatsagortsyan}
\email{k.hatsagortsyan@mpi-hd.mpg.de}
\affiliation{Max-Planck-Institut f\"{u}r Kernphysik, Saupfercheckweg 1, 69117 Heidelberg, Germany}
\author{Christoph H. Keitel}\affiliation{Max-Planck-Institut f\"{u}r Kernphysik, Saupfercheckweg 1, 69117 Heidelberg, Germany}

\date {\today}

\begin{abstract}

Generation of ultrarelativistic polarized positrons during interaction of an ultrarelativistic electron beam with a counterpropagating  two-color petawatt laser pulse is investigated theoretically.
Our Monte Carlo simulation based on a semi-classical model, incorporates photon emissions and pair productions, using spin-resolved quantum probabilities in the local constant field approximation, and describes the polarization of electrons and positrons for the pair production and photon emission processes, as well as the classical spin precession in-between. The main reason of the polarization is shown to be the spin-asymmetry of the pair production process in strong external fields, combined with the  asymmetry of the two-color laser field. Employing a feasible scenario, we show that highly polarized  positron beams, with a polarization degree of $\zeta\approx 60\%$, can be produced in a femtosecond time scale, with a small angular divergence, $\sim 74$ mrad, and high density $\sim 10^{14}$ cm$^{-3}$. The laser-driven positron source, along with  laser wakefield acceleration, may pave the way to small scale facilities for high energy physics studies.

\end{abstract}

\maketitle

Accelerated beams of polarized positrons have important applications in  high-energy physics, and solid-state physics. Polarized electrons and positrons are powerful probes for precise measurement of the nucleon spin structure by deep inelastic lepton scattering \cite{subashiev1998spin, voutier2014physics}, for  measurement of the electroweak mixing angle in M{\o}ller scattering \cite{anthony2005precision}, as well as for study of magnetic properties and inner structural defects of materials \cite{gidley1982polarized, krause1999positron}. The future International Linear Collider (ILC) is designed to bring revolutionary new insight into our understanding of the fundamental interactions, to test the Standard Model with unprecedented precision, and to search for new physics beyond it. A key element for maximizing the physics capability of ILC is the ability to provide intense (3$\times10^{10}$/bunch) highly polarized electron ( $>80\%$ polarization) and positron ($30\%\sim60\%$ polarization) sources \cite{moortgat2008polarized,vauth2016beam,scott2011demonstration}.

While there are several methods for electron polarization, for instance, illuminating a photocathode by  circularly polarized light  \cite{Pierce_1976}, or with radiative polarization in a storage ring via the Sokolov-Ternov effect  \cite{Sokolov_1964,Sokolov_1968,Baier_1967,Baier_1972}, however, it is much more demanding to obtain polarized positron beams. For the latter a two step method has been commonly applied. Firstly,  circularly polarized $\gamma$-photons are produced, which  are then converted to polarized positrons via pair production in a thin high-$Z$ target. Polarized $\gamma$-photons can be generated by Compton backscattering of circularly polarized laser light off ultrarelativistic electrons \cite{Sakai_2003,omori2006efficient}. With the same scheme, in the Stanford Linear Accelerator Center (SLAC) polarized positrons  have been successfully produced in an undulator-based experiment (E166),  where multi-MeV circularly polarized photons are produced by a 46.6-GeV electron beam passing through a helical undulator \cite{alexander2008observation,alexander2009undulator}.
%Another concept, Polarized Electrons for Polarized Positrons, has been proposed in the Thomas Jefferson National Accelerator Facility \cite{abbott2016production}, where the polarization of electrons is transferred to  bremsstrahlung photons and further to positrons via pair production, relaxing the requirement of initial electrons energy from GeV to MeV.
However, the high cost of providing
%high brightness electron beams,
suitable helical undulator and techniques to avoid destruction of target may limit the application of these schemes to a wide community. Moreover, the present technique available for ILC, is based on a 100 m long helical undulator, and allows positron sources with only up to $30\%$ polarization degree \cite{barish2006baseline}, and challenging technical upgrades are required to reach the ILC target parameters \cite{scott2011demonstration}.

Recent development of the strong laser technique, see e.g. \cite{Yanovsky_2008,Kessel_2018,ASTRA}, raised hopes to polarize electrons and positrons with laser fields. At first sight, the strong magnetic field of a laser pulse could induce a significant radiative polarization of electrons (positrons) interacting with it. However, unfortunately the laser field has an oscillating character, changing the field direction in subsequent field cycles, as a consequence in a monochromatic laser field the polarization of a particle due to nonlinear Compton scattering is vanishing \cite{Kotkin_2003,ivanov2004complete,karlovets2011radiative} and is small in a case of a short laser pulse \cite{seipt2018theory}. Recent progress in the theory shows that  significant polarization for electrons can be acquired in a  rotating electric field  \cite{del2017spin, del2018electron},  which models anti-nodes of a circularly polarized standing laser wave.
However the electrons cannot be trapped in anti-nodes of a circularly polarized standing wave \cite{bashinov2017particle}, in contrast to the  anomalous radiative trapping in the case of linear polarization  \cite{Gonoskov_2014}.
%note that the electrons can be trapped in anti-nodes only at linearly polarized standing wave  with extreme laser intensities $I\gtrsim 10^{26}$ W/cm$^2$ (in the anomalous radiative trapping regime) \cite{Gonoskov_2014}, and  no such regime exists in a circularly polarized standing wave \cite{bashinov2017particle}.
In another development, it has been shown recently \cite{li2018ultrarelativistic} that with a fine tuning of the ellipticity of the symmetric laser field, the splitting of the particle beam with respect to polarization can be achieved due to  spin-dependent radiation reaction. The same scheme is employed in the pair production regime to split the created positrons with respect to polarization \cite{JX}.

The petawatt laser technology \cite{danson2015petawatt, Vulcan,ELI,XCELS} enables
another potential method to produce polarized positrons  via the multiphoton Breit-Wheeler process \cite{Ritus_1985}, when
$e^-e^+$ pairs are produced due to a high-energy $\gamma$-photon interacting with a strong laser field, accompanied with multiple photon absorption from a strong laser field (firstly demonstrated in the famous  SLAC experiment E-144 \cite{burke1997positron}). Note that recently generation of unpolarized electron-positron  jets for laboratory astrophysics has been demonstrated experimentally in laser-solid interaction \cite{Chen_2009,Chen_2010,Liang_2015}, and  in laser-electron beam interaction \cite{Sarri_2015n}.

Generally, the radiative polarization of electrons (positrons) requires asymmetric laser fields, but what is remarkable, with a given asymmetric field the spin-dependent asymmetry  is stronger for the   pair production process than  for a photon emission. Consequently, the polarization of the electrons (positrons) during pair production process in that field generally will outstrip the radiative polarization. In previous studies \cite{tsai1993laser+,ivanov2005complete,jansen2016strong}, the polarization effects in the multiphoton Breit-Wheeler process have been considered in plane wave laser fields mostly of moderate intensity. It remained unclear  if the sizable polarization of positrons in  realistic strong laser fields is feasible.

%The spin effects in the multiphoton Breit-Wheeler process for a single photon emission (with polarized and unpolarized $\gamma$-photons) in different regimes in a monochromatic \cite{tsai1993laser+,ivanov2005complete},  and pulsed laser fields \cite{jansen2016strong}. in a constant crossed field \cite{Baier_1998,Shaisultanov} are investigated. In previous studies, the polarization effects have been considered in plane wave laser fields mostly of moderate intensity. What remained unclear is how the multiple photon emissions will affect the polarization degree of the created particles with ultrastrong laser field, and if the sizable polarization of positrons in  realistic laser fields is feasible.
 \begin{figure}
	\begin{center}
	\includegraphics[width=0.5\textwidth]{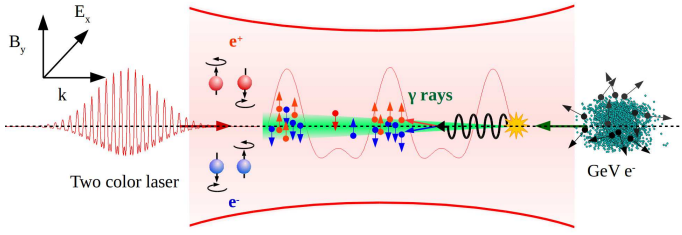}
	\end{center}
	\caption{Scheme of laser-based polarized positron beam production. An intense linearly polarized two-color laser pulse head-on collides with an unpolarized relativistic electron beam, resulting in  emission of $\gamma$-photons in forward direction, which decay into polarized $e^+$  and $e^-$, with spin parallel and anti-parallel to the laser's magnetic field direction, respectively, and with a small divergence angle of propagation.    }
	\label{Fig. 1}
\end{figure}

In this Letter, we investigate theoretically the feasibility of production of collimated and highly polarized ultrarelativistic positron beams during interaction of an ultrarelativistic electron beam with a counterpropagating two-color intense laser pulse in quantum radiation dominated regime, based on the setup in  Fig. \ref{Fig. 1}. During the interaction $\gamma$-photons via nonlinear Compton scattering are generated, which subsequently decay into electron-positron pairs in multiphoton Breit-Wheeler process. With Monte Carlo simulations we study the spin-resolved dynamics of electrons and positrons during pair production and photon emission processes, as well as during propagation in the laser field. In our scheme, positrons are mostly  polarized due to the intrinsic asymmetry of spin-resolved pair production probability in strong fields, combined with the asymmetry of the two-color laser field configuration. As a result, an ultrarelativistic positron beam with $60\%$ polarization, with uniformly distributed polarization within
%a 30 mrad
the divergence angle can be generated with realistic laser fields.

We employ a semi-classical method to describe the electron (positron) spin-resolved  dynamics in nonlinear Compton scattering \cite{Ridgers_2014,Elkina_2011,Green_2015}. The photon emissions and pair productions are simulated via spin-resolved quantum probabilities and the Monte Carlo algorithm, similar to \cite{li2018ultrarelativistic} (see also the more recent work \cite{geng2019radiative}). The employed spin resolved photon emission and pair production probabilities are derived with the QED operator method under the local constant field approximation \cite{Shaisultanov,Baier_1998,supplement}. The spin precession in the external field is described by Bargmann-Michel-Telegdi (BMT) equation \cite{Bargmann_1959,Walser_2002}. The accuracy of our Monte-Carlo code is confirmed by reproducing the results on the radiative polarization \cite{bauier1972radiative, jackson1976understanding, del2018electron, seipt2018theory,li2018ultrarelativistic} and pair production \cite{gonoskov2015extended, Sokolov:1986nk}, see \cite{supplement}.

 \begin{figure*}
	\begin{center}
	\includegraphics[width=0.8\textwidth]{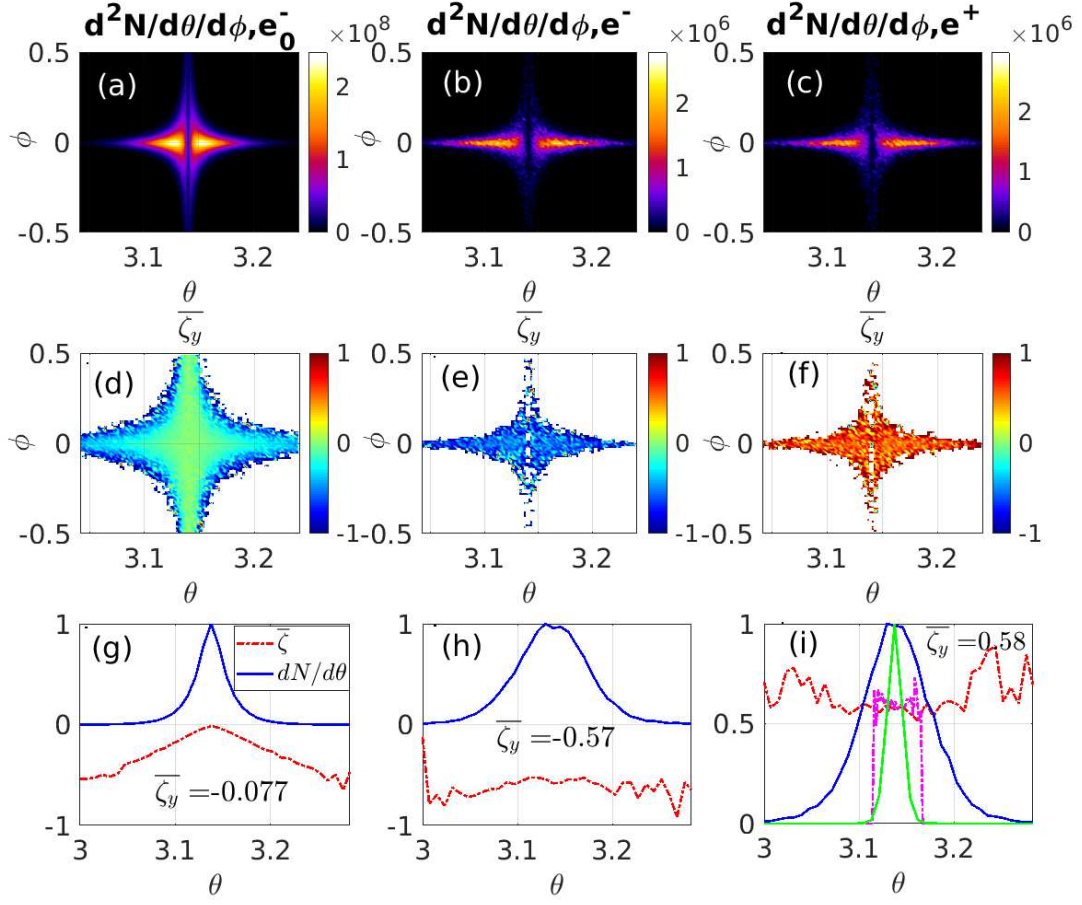}
	\end{center}
	\caption{(top) Angular distribution $d^2N/(d\theta d\phi)$, with the polar angle $\theta$ (rad), and the azimuthal angle $\phi$ (rad): (a) for seed electrons $e_0^-$, (b) for produced electrons $e^-$,  and (c) for positrons $e^+$. (middle row) The averaged spin distribution along the magnetic field direction $\overline{\zeta}_y$ vs $\theta$ and $\phi$: (d) for $e_0^-$,  (e) for $e^-$, and (f) for $e^+$. (bottom row) The particle number distribution $dN/d\theta$ (solid, blue), and the average spin component $\overline{\zeta_y}$  (dash-dot, red): (g) for $e_0^-$, (h) for $e^-$, and (i) for $e^+$; In (i) (green) $dN/d\theta$ and (magenta) $\overline{\zeta_y}$  are for $e^+$ at the production points [the beam average of $\overline{\zeta_y}$ are indicated on panels (g)-(i)];
at the production points $\overline{\zeta}_{y0}=0.6$);
	$\xi_0=100$, $ \varepsilon_0=2$ GeV. }
	\label{Fig. 2}
\end{figure*}

We consider interaction of an intense two-color linearly polarized laser pulse with a counterpropagating ultrarelativistic electron beam, as shown in Fig. \ref{Fig. 1}. The two-color field consists of two laser pulses of $\lambda_1=1\,\mu$m  and  $\lambda_2=0.5\,\mu$m wavelengths, and $\tau_{p1}=10T_1$ and $\tau_{p2}=20T_2$ pulse durations, respectively. Their peak intensities ($I_{0i}$, $i=1,2$) fulfill the ratio $R\equiv \xi_{1}/\xi_{2}=2$, with the dimensionless field parameter $\xi_i=|e|E_{0i}/(m\omega_i)=85.5\sqrt{\lambda_i\,[\mu\text{m}]^2 I_{0i}\,[10^{22}\text{W/cm}^2]}$, the laser field amplitude $E_{0i}$, and frequency $\omega_i$, the electron charge $e$, and mass $m$; the relativistic units with $\hbar=c=1$ are used throughout. The phase difference between the two color fields is chosen as $\Delta\phi=0$ to obtain the maximum field asymmetry.  The beam waist size is $5\lambda_1$ for both laser pulses. The cylinder electron beam has a longitudinal uniform distribution and transverse Gaussian distribution, with a standard deviation  $\sigma=0.4\lambda_1$, beam length $L_e=10\lambda_{1}$, the number of electrons in the bunch is $N_e=10^6$. The initial kinetic energy spread of electron bunch is $\Delta\varepsilon/\varepsilon_0=0.02$, and the angle divergence is $1$~mrad.

When choosing the laser intensity and the electron energy, we should take into account that the pair production is substantial when the quantum field parameter $\chi_{\gamma}\gtrsim 1$, where $\chi_{\gamma ,e}=|e|\hbar\sqrt{(F_{\mu\nu}p_{\gamma ,e}^\nu)^2}/m^3$, with $p_{\gamma ,e}$ being the 4-momentum of the $\gamma$-photon, and electron, respectively, and $F_{\mu\nu}$  the field tensor. One may assume that during nonlinear Compton scattering  photon and electron energies are of the same order, and estimate $\chi_{\gamma}\sim \chi_e$. In the counterpropagating setup $\chi_e\approx 2(\omega_i/m)\xi_i\varepsilon_0/m\approx 0.59\times\varepsilon_0\,[\text{GeV}]\sqrt{I[10^{22}\,\text{W/cm}^2]}$, with the electron initial energy $\varepsilon_0$.
%relativistic $\gamma$-factor.

As the presently available laser intensities do not exceed $10^{22}\,\text{W/cm}^2$ \cite{Yanovsky_2008}, and for the electron energies via laser wakefield acceleration the electron energy is less than 10~GeV \cite{leemans2014multi}, we choose as  typical parameters: the initial electron energy $\varepsilon_0=2$ GeV, and the laser  full  intensity parameter $\xi_0=\xi_{1}+\xi_{2}=100$. The produced positron beam parameters are summarized in Fig. \ref{Fig. 2}. The total number of produced positrons is $N_{+}\sim 2\times10^{4}$, which travel in forward direction, with  the full width at half maximum (FWHM) of the angular distribution $\theta_{+}\approx 74$ mrad, see Fig.~\ref{Fig. 2} (c), and (i). The averaged spin component of the produced positrons along magnetic field direction ($y$-axis) has a nearly uniform angular distribution ($\overline{\zeta_y}=0.58$) within FWHM, see Fig. \ref{Fig. 2} (f) and (i). The angle-independence  of the positron beam polarization renders further collimation feasible without polarization damage. The produced electrons have similar properties as positrons except the opposite sign for polarization, see Fig. \ref{Fig. 2} (b), (e), and (h).
% as the positrons polarized along the laser magnetic field, while the electrons oppositely.
The seed electrons have narrower angular distribution, as shown in Fig. \ref{Fig. 2} (a) and (g), but the polarization degree is significantly lower, $\overline{\zeta}_y=-0.077$, than that of the produced $e^-e^+$ pairs. The latter indicates that the polarization mechanisms for the seed electrons and created particles are different.

 \begin{figure*}
	\begin{center}
	\includegraphics[width=0.8\textwidth]{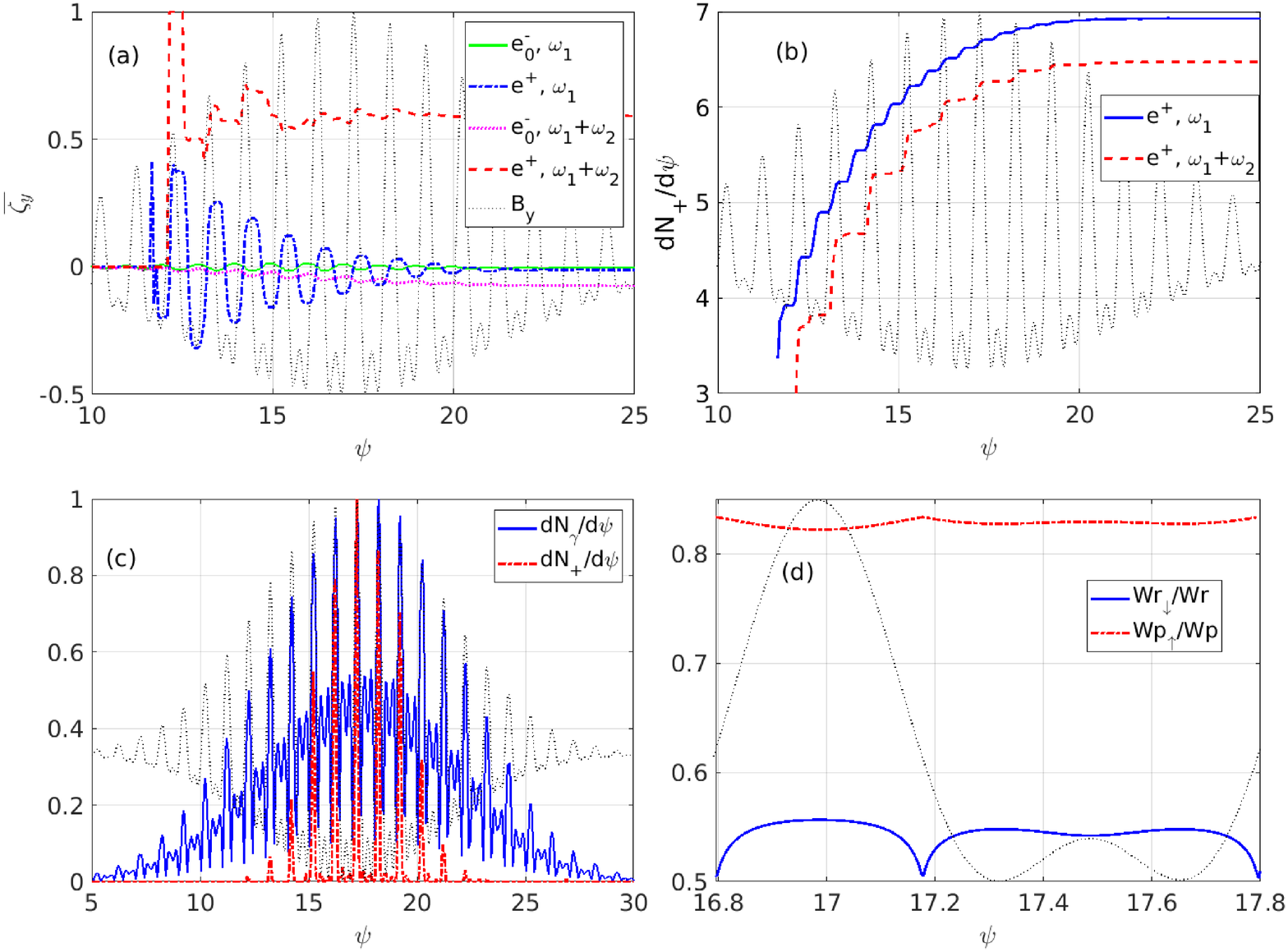}
	\end{center}
	\caption{(a) Averaged spin $\overline{\zeta}_y$ vs laser cycle $\psi$: in an one-color laser pulse, for seed electrons $e_0^-$ (solid, green), and positrons  $e^+$  (dash-dotted, blue); in  a two color laser pulse, $e_0^-$ (dotted, pink), and $e^+$  (dashed, red).  (b) Positron number $dN/d\psi$ vs $\psi$ for one- (solid, blue) and two-color (dashed, red)  laser pulses. (c) Photon number $dN_\gamma/d\psi$ emitted by $e_0^-$ (solid, blue) at emission points and positron number $dN_{+}/d\psi$  (dashed, red) at production points. (d) Ratio of electron radiation probability with spin anti-parallel to magnetic field direction (solid, blue), and ratio of pair production probability with the positron spin parallel to magnetic field direction (dash-dotted, red); $\xi_0=100$, $\varepsilon_0=2$ GeV; The faint dotted line is the two-color laser field.}
	\label{Fig. 3}
	\end{figure*}

To elucidate the reason for the significant polarization of positrons in the two-color laser field, we compare the spin dynamics in one- and two-color laser fields with $\xi_0=100$  in Fig.~\ref{Fig. 3}~(a). It can be seen that the polarization of seed electrons oscillates in the laser fields for both cases, but only in the two-color field seed electrons acquire a small $\overline{\zeta}_y$.  The reason is the almost complete symmetry in the laser magnetic field direction in the case of a single color field, and its asymmetry  in the two color field. After each photon emission the  electrons (positrons) are more likely to flip opposite (along) the magnetic field direction in the particle rest frame,
%(determined by a unit vector $\bm{n}=\bm{\beta}\times\bm{a}/|a|$, where $\bm{a}$ is acceleration of emitting particles \cite{gonoskov2015extended}),
which is approximately parallel (anti-parallel) to the magnetic field direction $B_y$ for  electrons (positrons) in the Lab-frame. This is because
%It is because when ultrarelativistic electrons head-on collide with a linearly polarized laser pulse and emit $\gamma$ rays in forward direction, the instantaneous quantization axis for both photon emission and pair production are roughly parallel with magnetic field direction (along $y-$axis), i.e. $B_y>0, \bm{n}\approx(0,1,0); B_y<0, \bm{n}\approx(0,-1,0)$. Since
the spin resolved emission probability is in favor of spin down for electrons. As shown in Fig. \ref{Fig. 3} (d), $W_{r\downarrow}/W_{r}>50\%$, where $W_{r}$ is the radiation total probability, and $W_{r\downarrow}$ is the probability with finial spin anti-parallel to the magnetic field.
%$W_{r\downarrow}=\sum_{\bm{\zeta_i}}W_r(\bm{\zeta_i},\bm{\zeta_f}=-\bm{n})$
% the  electrons are more likely to flip to $-\bm{n}$ after emission.
Therefore, $\overline{\zeta}_y$ of seed electrons increases during $B_y<0$ and decreases during $B_y>0$, see Fig. \ref{Fig. 3} (a). In the one-color laser pulse, $\overline{\zeta_y}$ oscillates symmetrically in consecutive laser cycles.
% with roughly a $\pi/2$ shift with respect to $B_y$ and goes to zeros outside of the laser, as shown in Fig. \ref{Fig. 3} (a).
While two-color field breaks this symmetric pattern in $\overline{\zeta}_y$ oscillation, more electrons emit photons and acquires $\zeta_y<0$ during $B_y>0$, which  results in a -7.7$\%$ polarization at the end of the interaction, see Fig.~\ref{Fig. 3}~(a).

\begin{figure}
	\begin{center}
	\includegraphics[width=0.5\textwidth]{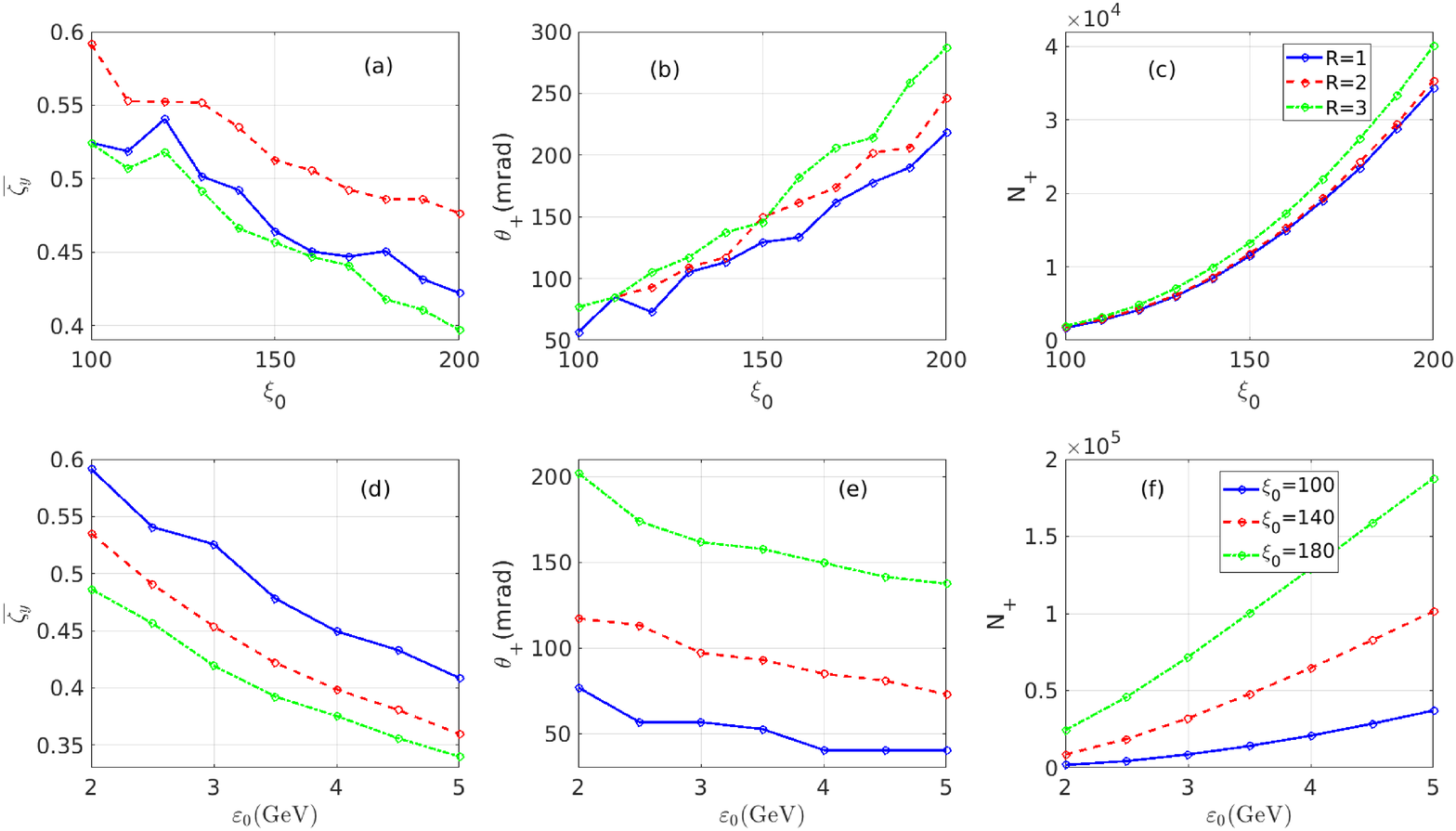}
	\end{center}
	\caption{  (a), (d) The averaged polarization $\overline{\zeta}_y$; (b), (e) FWHM of $\theta_{+}$, and (c), (f) positron number $N_{+}$. Top row: (solid, blue) $R=1$, (dashed, red) $R=2$,  and (dot-dahsed, green) $R=3$ vs laser peak intensity $\xi_0$ for $\varepsilon_0=2$ GeV. Bottom row: (solid, blue) $\xi_0=100$, (dashed, red) $\xi_0$=140, and (dot-dahsed, green) $\xi_0$=180, for $R=2$. The number of initial electrons is $N_{-}=2\times 10^5$.
}
	\label{Fig. 4}
\end{figure}

Unlike the seed electrons that are slightly polarized due to radiative polarization, the produced   positrons (electrons) are highly polarized $\overline{\zeta}_y\sim 60\%$, see Fig. \ref{Fig. 2} (i).
Although, the polarization of seed electrons and created pairs both benefit from the asymmetry of the two-color field configuration, however, the asymmetry has more significant impact on the polarization of the pairs than on the seed electrons. This is because seed electrons are polarized by radiative polarization due to photon emissions, while the pair polarization mainly comes from the creation process. There are two reasons for spin-asymmetry in the pair production process in an asymmetric field. First of all,  the spin related terms in the probabilities play a more important role in the case of pair production than in radiation, as shown in Fig. \ref{Fig. 3} (d). It is much more probable to have a positron along magnetic field during the pair production process (around $80 \%$), than due to photon emission (around $55 \%$). Secondly, the probability of the pair production has a stronger dependence on the laser intensity than the radiation probability. As shown in Fig. \ref{Fig. 3} (c), the pair production takes place mainly in the dominant half-cycles of the two-color field, and fully suppressed in the weak half-cycle,  while photons are emitted in both half-cycles with slightly different probabilities. The second is the main reason for highly polarized positrons in a two-color field.

Since positrons are more probable to be produced along the instantaneous laser magnetic field, in one-color symmetric laser field the averaged polarization is negligible after the interaction, as shown in Fig. \ref{Fig. 3} (a). While in the two-color asymmetric field a large polarization for positron beam is obtained, because positrons are mostly produced at $B_y>0$, see Fig. \ref{Fig. 3} (b) and (c), and the probability for these positrons to be polarized with $\zeta_y>0$ is very large, see Fig. \ref{Fig. 3} (d). Therefore, the positron density for $\zeta_y>0$ is far higher than that for $\zeta_y<0$.

We underline that the positron initial polarization degree during the production process further decreases due to photon emissions.  This is because the radiative polarization has less spin dependence than the pair production, according to Fig. \ref{Fig. 3} (d). Positrons produced with $\zeta_y>0$ have a chance to flip to $\zeta_y<0$ when emitting photons at $B_y<0$, which brings down the polarization.
Fortunately, the decrease of the polarization degree is not large from, the  $\overline{\zeta}_y\sim 60\%$ at the production point to the final $\overline{\zeta}_y\sim 58\%$, as it is shown in Fig.~\ref{Fig. 2} (i).

We have investigated the optimal conditions for positron polarization and the results are summarized in Fig.~\ref{Fig. 4}. Firstly, we varied the ratio ($R$) of laser intensities.
Fig. \ref{Fig. 4} (a) shows that $R=2$ is the optimal choice of relative laser intensities to obtain higher polarization. In fact, in this case the two-color field is most asymmetric, when the $B_y<0$ parts of the field reaches to minimum strength. Increasing laser intensity yields more pairs, see Fig. \ref{Fig. 3} (c), but reduces the averaged polarization and increases angular divergence, see Fig. \ref{Fig. 3} (a) and (b).  This is because higher laser intensity triggers more pair production in negative parts of laser fields, bringing in more positrons with $\zeta_y<0$. The similar scaling laws can also be found for electrons' initial energy, as shown in Fig. \ref{Fig. 4} (c)-(d). More energetic electrons give rise to more pairs and better emittance but compromise the polarization degree.
%For $\xi_0=180$ and $\varepsilon_0=5$GeV, $10^5$ positrons can be produced with polarization degree $30\%$.
Therefore, polarized positron beam production is a multidimensional problem with cross-talking parameters. If high polarization has priority over positron density, smaller laser intensity and initial electron energy is preferable to suppress contributions from $B_y<0$. If one aims for an intense $e^+$ beam with moderate polarization degree, higher laser intensity and more brightness of electrons are more preferable.\\

The number of initial electrons in our simulation is $10^6$, which yields a final electron number in the bunch $10^6$, and $10^4$ positrons in the bunch, with $8\%$ and $60\%$ polarization degree, respectively. The available laser wakefield acceleration technique can provide approximately $10^{10}$ electrons in the GeV regime \cite{leemans2014multi}, which will allow for $10^{8}$ polarized positrons in the bunch with $\zeta \approx 60\%$ degree of polarization ($10^{10}$ electrons per bunch with $\zeta \approx 8\%$). The latter parameters are sufficient to permit experiments for future colliders.

In summary, we have put forward a concept of two-color laser-based production of polarized positron beams. It employs an unpolarized electron driver and presently available ultrastrong lasers. In contrast to known radiative polarization, the polarization of positrons during their creation in the laser field is shown to be much more  sensitive to the field asymmetry, and allows for a high degree of polarization. This technique is not based on a large scale facility and opens access for polarized positron beams to a wide community.

%extends polarized positron capabilities from $50$ MeV to a few GeV,

P.-L.H. acknowledges the support from the China Scholarship Council (CSC).

\bibliography{1}

\end{document}